\documentclass[12pt]{article}

\usepackage{epsfig}
\usepackage{amsmath,amsfonts,amssymb}
\def\beq{\begin{equation}}
\def\eeq#1{\label{#1}\end{equation}}
\def\eeqn{\end{equation}}

\def\beqa{\begin{eqnarray}}
\def\eeqa#1{\label{#1}\end{eqnarray}}
\def\eeqan{\end{eqnarray}}

\let\bar=\overbar

\def\bra#1{\left\langle{ #1} \right|}
\def\ket#1{\left| {#1} \right\rangle}

\def\Dslash{\not{\hbox{\kern-4pt $D$}}}
\def\dslash{\not{\hbox{\kern-2pt $\del$}}}

\def\msb{{\bar{\ssstyle M \kern -1pt S}}}

\def\Title#1{\begin{center} {\Large {\bf #1} } \end{center}}

\begin{document}

\Title{Lattice Calculations of $B\to K/K^*l^+l^-$ form factors}

\bigskip\bigskip


\begin{raggedright}  
{\it Ran Zhou\index{Zhou, R.}\\
Department of Physics\\
Indiana University\\
bloomington, IN, USA}
\bigskip\bigskip
\end{raggedright}

\section{Introduction}

The exclusive rare semileptonic decays $B\to Kl^+l^−$, $B \to K^∗l^+l^−$, and the 
corresponding quark level process $b \to sl^+l^−$ are  mediated by flavor changing 
neutral currents (FCNC). In the standard model such processes arise at one-loop and 
are therefore suppressed. Hence, these decays are good candidates to search for new 
physics beyond the SM. 
Experimental measurements of $B\to K/K^*l^+l^-$ decays have been performed by the 
BABAR, Belle, and CDF Collaborations~\cite{Aubert:2008ps,Wei:2009zv,Aaltonen:2011cn}. Most recently, 
new results have been reported by LHCb~\cite{Aaij:2011aa,LHCb:2012ab,Aaij:2012cq}. We expect to 
see increasingly accurate results from both LHCb and from planned high-intensity 
$B$ factories~\cite{Hewett:2012ns}.

To find evidence of new physics, it is necessary to compare the experimental results with theoretical 
predictions from the SM. The current experimental results are consistent with the SM predictions,
but when the experimental error decreases, we need more accurate theoretical predictions. 
The accuracy of the theoretical prediction in $B\to K/K^*l^+l^-$ is limited by the error of the hadronic matrix 
elements $\bra{B}\hat{O}\ket{K/K^*}$, which are parameterized by form factors. Some earlier 
theoretical calculations were based on the form factors calculated from Light Cone Sum Rules (LCSR). LCSR computes
the form factors at low $q^2$, where $q^2$ is the outgoing dilepton invariant mass 
squared. The form factors at large $q^2$ are extrapolated from low $q^2$ 
results, which ends in a large error. Lattice QCD can calculate form factors directly at large 
$q^2$ from first principles. Some earlier lattice QCD works calculated $B\to K/K^*l^+l^-$ form 
factors within quenched 
approximation~\cite{Becirevic:2006nm, Del.Debbio:1997kr, Abada:1995fa, Bhattacharya:1994fp, Bowler:1993rz, Bernard:1993yt, AlHaydari:2009zr}. These works obtained form factors at large $q^2$ within 15\%-20\% accuracy.
Moreover, the quenched calculations have the systematic error from lack of sea quark, which 
is difficult to remove. 
Modern lattice QCD simulations include realistic sea quark effects, hence removing this systematic error. 
Recently, three lattice collaborations (FNAL/MILC, HPQCD, and Cambridge/W\&M/Edinburgh)
have started calculations of $B\to K/K^*l^+l^-$ form factors based on the 2+1 flavor ensembles generated by 
the MILC collaboration~\cite{Bernard:2001av}. 
The FNAL/MILC collaborations and HPQCD collaboration are working on the $B\to Kl^+l^-$ 
decay~\cite{Zhou:2011be,Zhou:2012sn,Bouchard:2012tb}. The 
Cambridge/W\&M/Edinburgh collaboration are working on both of these two decays~\cite{Liu:2009dj,Liu:2011rra}. 

Current methods in lattice QCD allow for calculations of  $B \to K l^+l^-$ with complete control 
over all sources of systematic error. This is not the case for  $B \to K^*l^+l^-$, however, due 
to the fact that the $K^*$ meson is unstable, and needs to be treated as a resonance. Hence 
current lattice QCD calculations of this process contain an additional systematic error that 
is difficult to quantify~\cite{Liu:2009dj}. 

\section{Lattice formalism}
The three new lattice QCD calculations use $N_f$=2+1 flavor gauge configurations generated
by the MILC Collaboration~\cite{Bazavov:2009bb}.  
The MILC collaboration used the tree-level improved L\"{u}scher-Wise action for the gauge fields
and asqtad-improved staggered action for light sea quarks.
These improvements suppress the lattice artifacts to the order of ${\cal{O}}(\alpha_s a^2)$ 
for the gluon field and ${\cal{O}}(\alpha_s a^2)$, ${\cal{O}}(a^4)$ for the fermion 
field~\cite{Bazavov:2009bb}, where $a$ denotes the lattice spacing. 
These ensembles have four lattice spacings which are about 0.12fm, 
0.09fm, 0.06fm, and 0.045fm. 
The FNAL/MILC collaborations employ ensembles on these four lattice spacings.
The HPQCD collaboration and  Cambridge/W\&M/Edinburgh collaboration employ the ensembles 
with lattice spacings at $a \approx 0.12$fm and $a \approx 0.09$fm.
Although these three groups use the similar gauge ensembles, they employ different actions for valence $b$ and $s$
quarks. Both FNAL/MILC and Cambridge/W\&M/Edinburgh group use the asqtad-improved staggered action for 
the valence strange quark. HPQCD uses HISQ action for valence strange quark~\cite{Follana:2006rc}. 
The HISQ action is more improved than the asqtad action~\cite{Follana:2006rc,Bazavov:2009bb}. 
For the heavy quark, 
the FNAL/MILC collaborations use the Sheikholeslami-Wohlert (SW)  action~\cite{Sheikholeslami:1985ij} 
with the Fermilab interpretation for the $b$ quark~\cite{ElKhadra:1996mp}. This action can be systematically improved 
and FNAL/MILC collaborations tune the $b$ quark hopping parameter $\kappa$ and clover coefficient $c_{\rm SW}$ to 
remove the discretization errors through next-to-leading order (${\mathcal{O}}(1/m_b)$)~\cite{Kronfeld:2000ck,Oktay:2008ex}. 
The HPQCD collaboration and Cambridge/W\&M/Edinburgh collaboration
use (moving)-NRQCD method~\cite{Lepage:1992tx,Horgan:2009ti} for the $b$ quark. The NRQCD method expands
the relativistic QCD action by the order of $v_b$ which is the velocity of the $b$ quark. The heavy quark action
is tuned to include the ${\mathcal{O}}(\Lambda_{\rm QCD}^2/m_b^2)$ corrections.
Because these three  groups are working on similar quantities with different discretization methods,
it provides a good cross check for their results.

\section{The lattice QCD calculation on form factors}
The theoretical description of the $B \to K/K^*l^+l^-$ process is based on
the Operator Production Expansion (OPE). The low energy effective Hamiltonian 
for the $b\to sl^+l^-$ transition is~\cite{Beneke:2001at, Beneke:2004dp, Chetyrkin:1996vx, Bobeth:2010wg} 
\begin{align}
{\cal H}_{eff}&=-\frac{4G_F}{\sqrt{2}}V_{tb}V_{ts}^*\sum_i {\cal C}_i(\mu){\cal O}_i(\mu) \,
\end{align}
where ${\cal O}_i$s are four-fermion operators of dimension six. ${\cal C}_i$s are the
corresponding Wilson coefficients. Most of the SM contribution is from the operators 
${\cal O}_{7,9,10}$. 
The operator ${\cal O}_7$ is a photon dipole operator and ${\cal O}_{9,10}$ are semileptonic operators. 
Theoretical predictions are calculated from ${\cal H}_{eff}$ and contain hadron matrix elements that 
are parametrized by form factors. For $B\to Kl^+l^-$ decay, there are three form factors
$f_+$, $f_0$, and $f_T$
\begin{eqnarray}
\bra{K} i\bar{s}\gamma^\mu b \ket{B}&=&f_+(q^2)\left(p_B^\mu+p_K^\mu-\frac{M_B^2-M_K^2}{q^2}q^\mu \right)
+f_0(q^2)\frac{M_B^2-M_K^2}{q^2}q^\mu, \label{eq:def.f+f0}\\
\bra{K}i\bar{s}\sigma^{\mu\nu} b\ket{B}&=&\frac{2f_T(q^2)}{M_B+M_K} (p^\mu k^\nu-p^\nu k^\nu)q_\nu, \label{eq:def.fT}
\end{eqnarray}
where $p_B$ and $p_K$ are the $B$ meson and kaon momenta, respectively. 
FNAL/MILC collaborations and HPQCD calculate these two matrix elements in the $B$ meson rest frame. 
The $q^2$ becomes $M_B^2+M_K^2-2 M_B E_K$ in this reference frame. The matrix elements are reparametrized as
\begin{eqnarray}
\bra{K} i\bar{s}\gamma^\mu b \ket{B}&=&\sqrt{2M_B}\left [ v^\mu f_\parallel(E_K)+p_\perp^\mu f_\perp(E_K)\right ],
\end{eqnarray}
where $v^\mu=p_B^\mu/M_B$ is the four-velocity of the $B$ meson and $p_\perp^\mu=p_K^\mu-(p_K\cdot v)v^\mu$. 
The form factors $f_\parallel$ and $f_\perp$ are solved
from the temporal and spatial components of the matrix element  of the vector current.  
Finally, the form factors ($f_+$, $f_0$) are reconstructed 
from $f_\parallel$ and $f_\perp$ by
\begin{eqnarray}
f_+ & = & \frac{1}{\sqrt{2M_B}}\left [ f_\parallel+(M_B-E_K)f_\perp \right],\\
f_0 & = & \frac{\sqrt{2M_B}}{M_B^2-M_K^2}\left [ (M_B-E_K)f_\parallel+(E_K^2-M_K^2)f_\perp \right] .
\end{eqnarray}
A similar method can be applied to $f_T$: 
\begin{eqnarray}
f_T&=&\frac{M_B+M_K}{\sqrt{2M_B}}\frac{\bra{K} ib\sigma^{0i} s\ket{B}}{\sqrt{2M_B}p_K^i} .
\end{eqnarray}

For $B\to K^*l^+l^-$ decay, there are four hadronic matrix elements related to  the theoretical predictions. They are
\begin{eqnarray}
\bra{K^*(k,\epsilon)}\bar{s}\gamma^\mu b\ket{B(p)}&=& \frac{2iV(q^2)}{M_B+M_{K^*}}\epsilon^{\mu\nu\rho\sigma}
\epsilon^*_\nu k_\rho p^\prime_\sigma\\
\bra{K^*(k,\epsilon)}\bar{s}\gamma^\mu \gamma_5 b\ket{B(p)}&=&2M_{K^*}A_0(q^2)\frac{\epsilon^*\cdot q}{q^2}q^\mu+
(M_B+M_{K^*})A_1(q^2)(\epsilon^{*\mu}-\frac{\epsilon^*\cdot q}{q^2}q^\mu) \nonumber \\
&&-A_2(q^2)\frac{\epsilon^*\cdot q}{M_B+M_{K^*}}(p^\mu+k^\mu-\frac{M_B^2-M_{K^*}^2}{q^2}q^\mu) \\
q^\nu\bra{K^*(k,\epsilon)}\bar{s}\sigma_{\mu\nu} b\ket{B(p)}&=& 4T_1(q^2)\epsilon_{\mu\nu\kappa\rho}
\epsilon^*_\rho p_\kappa k_\sigma\\
q^\nu\bra{K^*(k,\epsilon)}\bar{s}\sigma_{\mu\nu} \gamma_5 b\ket{B(p)}&=&2iT_2(q^2)
[\epsilon^*_\mu(M_B^2-M_{K^*}^2)-(\epsilon^*\cdot q)(p+k)_\mu] \nonumber \\
&&2iT_3(q^2)(\epsilon^*\cdot q)[q_\mu-\frac{q^2}{M_B^2-M_{K^*}^2}(p+k)_\mu]
\end{eqnarray}
Like the $f_{+,0,T}$ case, lattice QCD calculates the form factors by calculating a particular
component of the matrix elements. Details on $B\to K^*l^+l^-$ form factors are in 
Ref.~\cite{Liu:2009dj,Liu:2011rra}.

\section{Preliminary results on $B\to K/K^*l^+l^-$ form factors}
\subsection{Extract form factors on the lattices}
First of all, we extract the meson masses and energies 
from the two-point correlation functions measured on the gauge configurations:
\begin{eqnarray}
C_2(t;\vec p)&=&\sum_x \langle {\cal{O}}_P(\vec{x},t) {\cal{O}}_P^+(\vec{0},0) \rangle e^{-\vec{p}\cdot \vec{x}} 
=\sum_m (-1)^{mt}\frac{|\bra{0}{\cal O}_P\ket{P}|^2}{2E_P^{(m)}} e^{-E_P^{(m)} t} . 
\label{eq:2pt}
\end{eqnarray}
where $P$ represents the meson we want to study and ${\cal{O}}_P$ is the interpolating operator. 
If we insert a complete set of states, the two-point correlation functions are decomposed into 
the contributions from different energy levels. 
The $m$ labels the complete set of states that contribute to the sum. The factor $(-1)^{mt}$ arises 
only with staggered valence quarks. 
We are interested in the ground state energy and the excited states contributions can be 
safely neglected at large enough $t$. 
The FNAL/MILC collaborations use different smearing functions for $B$ and $K$ mesons to improve the accuracy of the
results~\cite{Bailey:2008wp,Zhou:2012sn}. 
The Cambridge/W\&M/Edinburgh group applies the all-to-all propagator technique~\cite{Foley:2005ac} to improve the 
signal-to-noise ratio. 
The ground state mass and energy in the current calculations are well-determined and have 
sub-percent statistical errors. 

Lattice QCD extracts hadronic matrix elements from the three-points correlation function.
For example, to determine $f_+$ and $f_0$ in $B\to Kl^+l^-$ decay, we 
measure $C_{3,\mu}(t,T; \vec{p}_K)$ which is defined as
\begin{eqnarray}
  C_{3,\mu}(t,T; \vec{p}_K) & = & \sum_{\vec{x},\vec{y}} e^{i \vec{p}_K \cdot \vec{y}} \langle {\cal O}_K (0,\vec{0})\, V_\mu (t,\vec{y})\, {\cal O}^\dagger_B (T,\vec{x}) \rangle 
\label{eq:R3mu}
\end{eqnarray}
where $V_\mu$=$i \bar{s} \gamma_\mu b$. $T$ denotes the location of the sink operator. 
Similar to the two-point correlation function, if we insert a complete set of states to 
the three-points correlation function $C_{3,\mu}$, it is decomposed into a sum of energy levels as 
\begin{align}
C_{3,\mu}(t,T;\vec p)&=\sum_{m,n} (-1)^{mt} (-1)^{n(T-t)}A^{mn}_{\mu}e^{-E_K^{(m)}t}e^{-M_B^{(n)}(T-t)} \ ,
\end{align}
where 
\begin{align}
A_\mu^{mn}&=\frac{\bra{0}{\cal O}_K\ket{K^{(m)}}}{2E_K^{(m)}} \bra{K^{(m)}}V_\mu\ket{B^{(n)}}\frac{\bra{B^{(n)}}{\cal O}_B\ket{0}}{2M_B^{(n)}} \ .
\end{align}
The $A^{00}_\mu$ contains the matrix elements we want. The three-points correlation 
function also has the contributions from
the excited states and opposite-parity states. Different methods are used to extract $A^{00}_\mu$. 
The FNAL/MILC collaborations apply an iterative averaging trick~\cite{Bailey:2008wp} to suppress the 
oscillating states contributions. The HPQCD and Cambridge/W\&M/Edinburgh group fit the two-point 
and three-point correlation functions simultaneously to extract $A^{00}_\mu$. They use the 
constrained fit technique~\cite{Lepage:2001ym,Hornbostel:2011hu}, which helps to resolve the information 
from excited states. 

\subsection{Chiral-continuum extrapolations and $z$-expansion fit}
\begin{figure}[ht]
\includegraphics[scale=0.85]{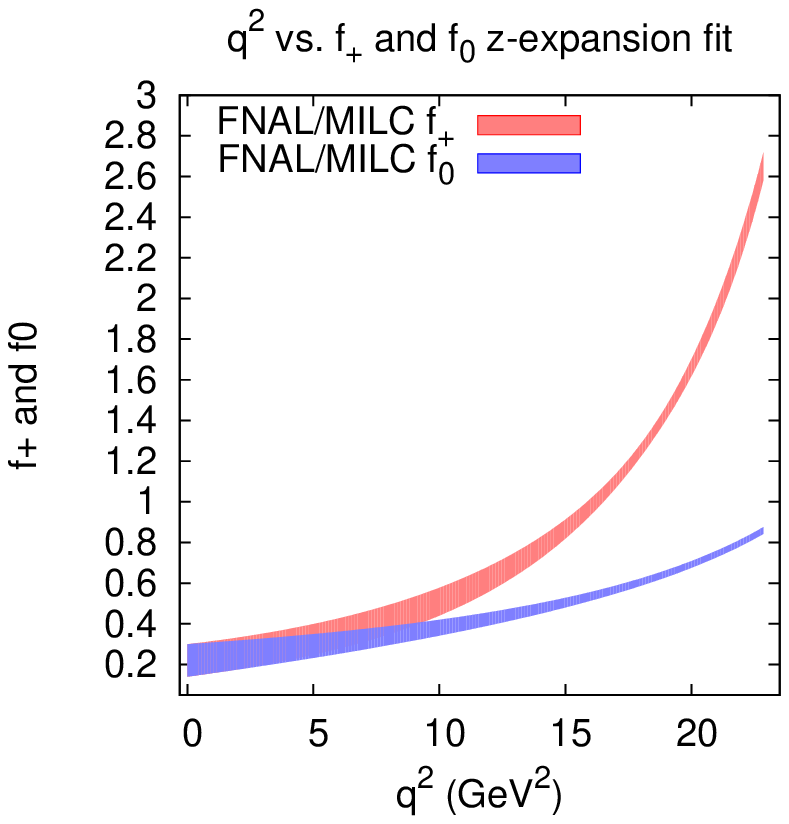}
\includegraphics[scale=0.85]{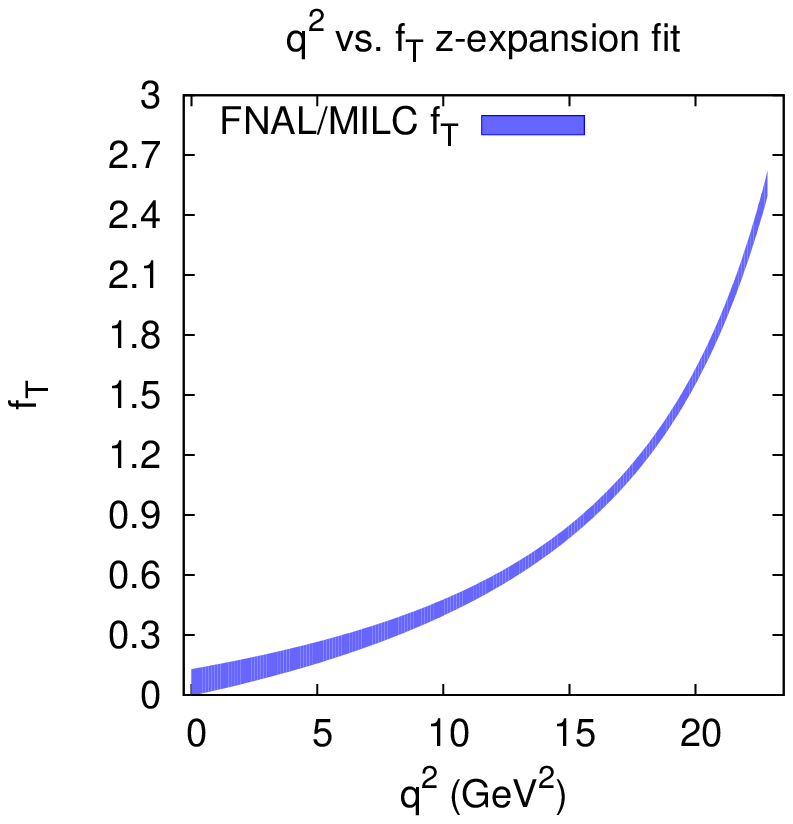}
\caption{Preliminary result of $f_{+,0}$ (left) and $f_T$ (right) on the whole $q^2$ range 
from FNAL/MILC collaborations. They obtain the form factors directly in the range 
$q^2 \gtrsim 15 {\textrm {GeV}}^2$ , and 
extrapolate them using the z-expansion into the region of $q^2 \lesssim  15 {\textrm {GeV}}^2$. }
\label{fig:fnal_milc_b2k}
\end{figure}

Although the strange quark mass ($m_s$) 
used in numerical simulations is typically close to its physical value, the light 
($u$, $d$) quark masses in current simulations are usually larger than their 
physical values. Very recent new simulations include light quarks with masses 
at their physical values~\cite{:2012uw,Durr:2010vn,Durr:2010aw}, but they have not yet been 
used to analyze $B$ meson decays. 
The FNAL/MILC collaborations perform combined continuum-chiral extrapolations using 
HMS$\chi$PT. $SU(3)$ HMS$\chi$PT was tested in 
$B$ and $D$ semileptonic decays~\cite{Bailey:2008wp,Bailey:2011bh}. 
Some preliminary results from FNAL/MILC collaborations suggest the $SU(2)$ HMS$\chi$PT might be
a better effective theory in the $B\to Kl^+l^-$ process. 
The continuum form factors from lattice data are at the small $E_K$ regime, 
because the discretization error becomes larger at large $E_K$ in lattice calculations. 
Moreover, the correlation functions become noisier and HMS$\chi$PT is not reliable at large $E_K$. 
To have the form factors on the whole $q^2$ range, the FNAL/MILC collaborations use the $z$-expansion 
to extrapolate the lattice results to low $q^2$ range. The $z$-expansion fit is a 
model-independent parametrization of form factors on the whole $q^2$ range. It maps the variable of 
$q^2$ to $z$:
\begin{align}
z(q^2,t_0)&=\frac{\sqrt{t_+-q^2}-\sqrt{t_+-t_0}}{\sqrt{t_+-q^2}+\sqrt{t_+-t_0}} ,
\end{align}
where $t_\pm=(M_B\pm M_{K/K^*})^2$. $t_0$ is selected to keep the absolute value of $z$ smaller than 1. The form factors 
are then parametrized as:
\begin{align}
f(q^2)&=\frac{1}{B(q^2)\phi(q^2, t_0)}\sum_{k=0}^\infty a_k z^k , \label{eq:zexp}
\end{align}
where $B(q^2)=z(q^2,m_R^2)$ is called the Blaschke factor. 
$m_R$ denotes the location of the pole in form factors. 
$\phi(q^2, t_0)$ is called the outer function. 
In fits of lattice (or experimental) data, the $z$-expansion is truncated at 
some finite order. Different choices for $B$ and $\phi$ yield different expansion 
coefficients. Generally, $\phi$ can be chosen to keep the $a_i$ small and 
hence the truncation error is well controlled. In addition, one can obtain bounds 
on the $a_i$ based on unitarity and heavy quark power counting. Hence the 
$z$-expansion provides us with a systematically improvable description of 
the $q^2$ dependence of the form factors. 
The preliminary results on form factors in $B\to Kl^+l^-$ from the FNAL/MILC collaborations
are summarized in Fig.~\ref{fig:fnal_milc_b2k}. The continuum form factors at large 
$q^2$ ($q^2 \gtrsim 15 {\textrm {GeV}}^2$) are obtained 
from chiral-continuum extrapolations with $SU(2)$ HMS$\chi$PT. The form factors at low $q^2$ 
($q^2 \lesssim  15 {\textrm {GeV}}^2$) are from 
the $z$-expansion fit. The systematic errors from chiral-continuum extrapolations, heavy and light 
quark discretization, renormalization factors, scale determination, light quark mass determination, 
and finite volume effect are included. The total error at large $q^2$ is about 5\%, which 
is more accurate than the previous quenched results. 

\begin{figure}[ht]
\includegraphics[scale=0.85]{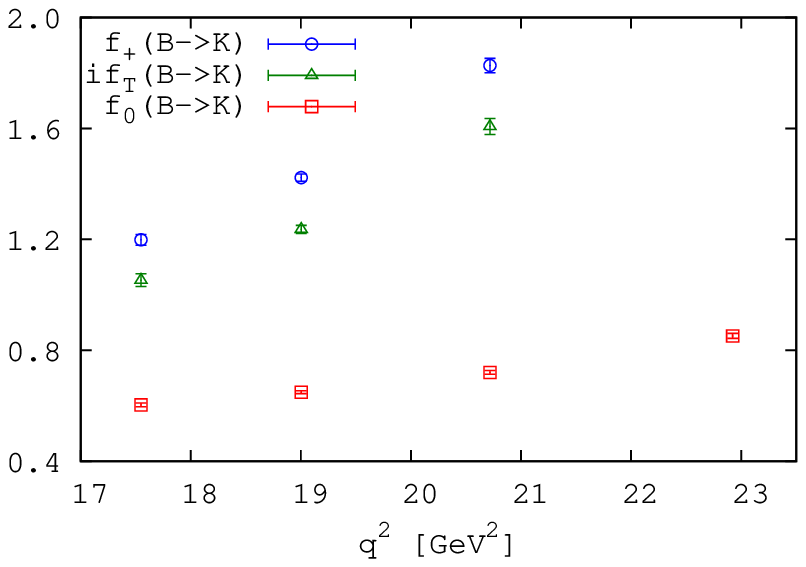}
\includegraphics[scale=0.35]{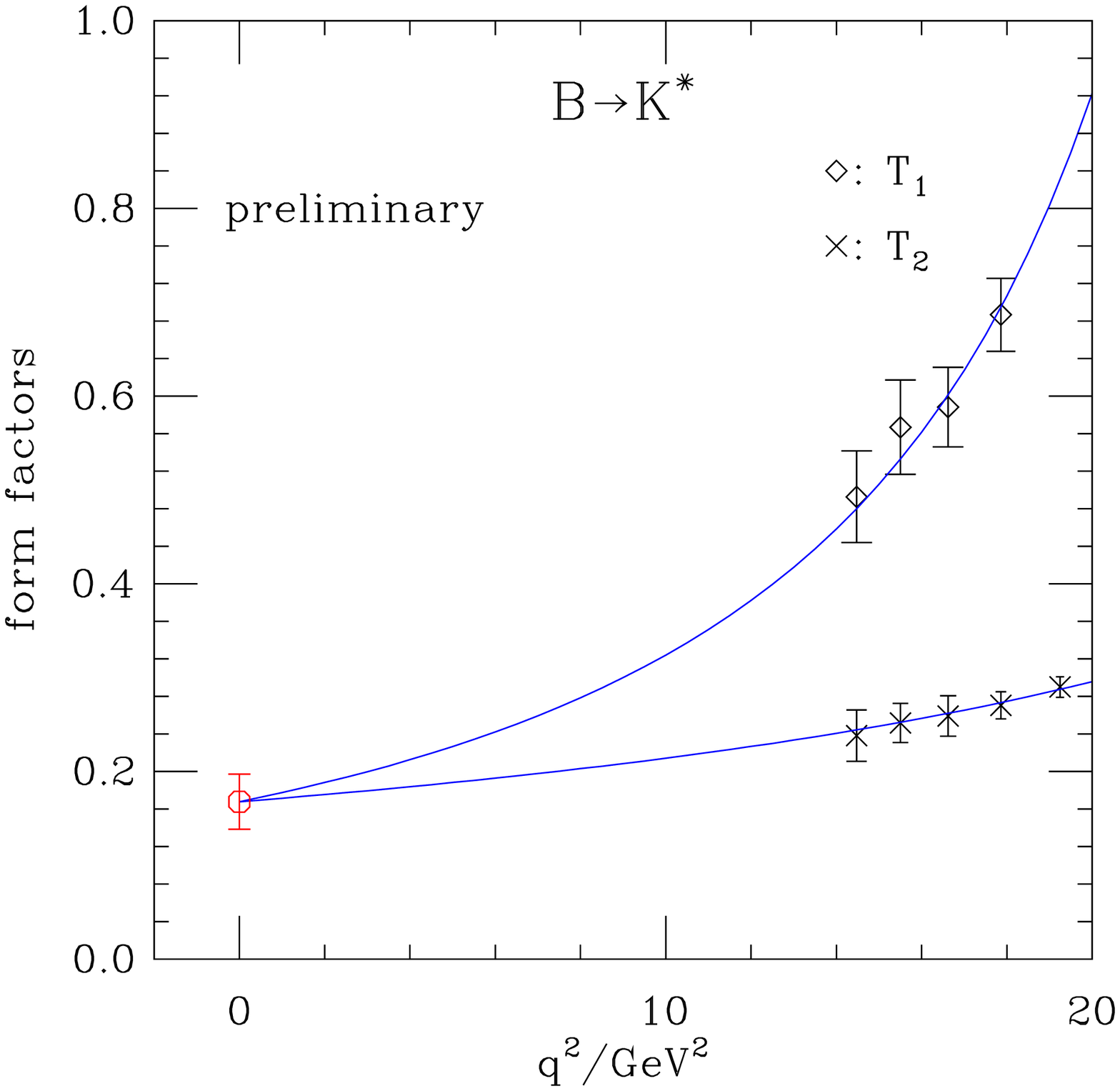}
\caption{The left panel is from C. Bouchard in the HPQCD collaboration. It is the preliminary result of 
$B\to Kl^+l^-$ $f_{+,0,T}$ measured on the lattices. Only statistical error is included and it is
about 1\% accuracy. The right panel is quoted from Ref.~\cite{Liu:2011rra}. It is the preliminary 
result of the $T_{1,2}$ in $B\to K^*l^+l^-$ process from Cambridge/W\&M/Edinburgh collaborations. }
\label{fig:hpqcd.cambridge}
\end{figure}

The HPQCD collaboration and Cambridge/W\&M/Edinburgh collaborations use modified $z$-expansion 
in the extrapolation of lattice form factors to continuum. As described in the last paragraph, 
the $z$-expansion fit provides a way to parametrize form factors on the whole $q^2$ range. 
The HPQCD collaboration 
uses a modified version of $z$-expansion, where heuristic discretization and light 
quark mass dependent terms are added in order  to perform combined chiral, continuum 
and shape fits~\cite{Na:2010uf}. 
The HPQCD collaboration applied it on the $D$ semileptonic decays~\cite{Na:2010uf} and 
is planning to use this method for 
 the $B\to Kl^+l^-$ process. The Cambridge/W\&M/Edinburgh collaborations also 
use the same method for  the extrapolations in $B\to K/K^*l^+l^-$ 
decays. The preliminary 
results from these two groups are summarized in Fig.~\ref{fig:hpqcd.cambridge}. The left panel is an example for 
$f_{+,0,T}$ in the $B\to Kl^+l^-$ process from the HPQCD collaboration. The HPQCD collaboration measures form factors with 
1\% accuracy at large $q^2$ on the $a\approx 0.12$fm MILC lattice ensembles (only statistical error is included 
here.). More measurements will be done in the future on the 
 $a\approx 0.09$fm MILC ensembles. Similarly, the right panel is the preliminary result 
of the $T_{1,2}$ in $B\to K^*l^+l^-$ process 
from the Cambridge/W\&M/Edinburgh collaborations. 

\section{Summary} 
Lattice QCD can calculate the form factors in $B\to K/K^*l^+l^-$ decays from first principles. 
Previous quenched calculations obtained the form factors with an uncertainty of 15\%-20\% 
and over a range of $q^2$ that was limited to large values.
The FNAL/MILC collaborations,
HPQCD collaboration, and Cambridge/W\&M/Edinburgh collaborations are 
working on the new calculations with dynamical QCD configurations, which include the sea quark effect. 
The analysis techniques like constrained fit, 
$SU(2)$ HMS$\chi$PT, and modified $z$-expansion are used in these new calculations. 
The preliminary $B\to Kl^+l^-$ form factors from FNAL/MILC collaborations 
have a total (combined statistical plus systematic) accuracy of about 5\%
at large $q^2$. Their final results will include a detailed and complete 
systematic error budget.
We can expect increasingly more precise and interesting results from lattice QCD 
calculation of these rare decays in the future. Sufficiently accurate theoretical 
predictions of the form factors are essential for making maximal use of experimental 
measurements, and may yield interesting constraints on new physics in the future.

\section{Acknowledgements}
The author thanks the usefully discussions with the members in the FNAL/MILC 
collaborations. The author also want to thank C. Bouchard and M. Wingate provide the progress of 
the works from the HPQCD collaboration and  Cambridge/W\&M/Edinburgh collaborations.


\begin{thebibliography}{99}

\bibitem{Aubert:2008ps} 
  B.~Aubert {\it et al.}  [BABAR Collaboration],
  Phys.\ Rev.\ Lett.\  {\bf 102}, 091803 (2009)
  [arXiv:0807.4119 [hep-ex]].

\bibitem{Wei:2009zv} 
  J.~-T.~Wei {\it et al.}  [BELLE Collaboration],
  Phys.\ Rev.\ Lett.\  {\bf 103}, 171801 (2009)
  [arXiv:0904.0770 [hep-ex]].

\bibitem{Aaltonen:2011cn} 
  T.~Aaltonen {\it et al.}  [CDF Collaboration],
  Phys.\ Rev.\ Lett.\  {\bf 106}, 161801 (2011)
  [arXiv:1101.1028 [hep-ex]].

\bibitem{Aaij:2011aa} 
  RAaij {\it et al.}  [LHCb Collaboration],
  Phys.\ Rev.\ Lett.\  {\bf 108}, 181806 (2012)
  [arXiv:1112.3515 [hep-ex]].

\bibitem{LHCb:2012ab} 
  RAaij {\it et al.}  [LHCb Collaboration],
  Phys.\ Rev.\ D {\bf 85}, 112013 (2012)
  [arXiv:1202.6267 [hep-ex]].

\bibitem{Aaij:2012cq} 
  RAaij {\it et al.}  [LHCb Collaboration],
  JHEP {\bf 1207}, 133 (2012)
  [arXiv:1205.3422 [hep-ex]].

\bibitem{Hewett:2012ns} 
  J.~L.~Hewett, H.~Weerts, R.~Brock, J.~N.~Butler, B.~C.~K.~Casey, J.~Collar, A.~de Govea and R.~Essig {\it et al.},
  arXiv:1205.2671 [hep-ex].

\bibitem{Becirevic:2006nm} 
  D.~Becirevic, V.~Lubicz and F.~Mescia,
  Nucl.\ Phys.\ B {\bf 769}, 31 (2007)
  [hep-ph/0611295].

\bibitem{Del.Debbio:1997kr} 
  L.~Del Debbio {\it et al.}  [UKQCD Collaboration],
  Phys.\ Lett.\ B {\bf 416}, 392 (1998)
  [hep-lat/9708008].

\bibitem{Abada:1995fa} 
  A.~Abada {\it et al.}  [APE Collaboration],
  Phys.\ Lett.\ B {\bf 365}, 275 (1996)
  [hep-lat/9503020].

\bibitem{Bhattacharya:1994fp} 
  T.~Bhattacharya and R.~Gupta,
  Nucl.\ Phys.\ Proc.\ Suppl.\  {\bf 42}, 935 (1995)
  [hep-lat/9501016].

\bibitem{Bowler:1993rz} 
  K.~C.~Bowler {\it et al.}  [UKQCD Collaboration],
  Phys.\ Rev.\ Lett.\  {\bf 72}, 1398 (1994)
  [hep-lat/9311004].

\bibitem{Bernard:1993yt} 
  C.~W.~Bernard, P.~Hsieh and A.~Soni,
  Phys.\ Rev.\ Lett.\  {\bf 72}, 1402 (1994)
  [hep-lat/9311010].

\bibitem{AlHaydari:2009zr} 
  A.~Al-Haydari {\it et al.}  [QCDSF Collaboration],
  Eur.\ Phys.\ J.\ A {\bf 43}, 107 (2010)
  [arXiv:0903.1664 [hep-lat]].

\bibitem{Bernard:2001av} 
  C.~W.~Bernard, T.~Burch, K.~Orginos, D.~Toussaint, T.~A.~DeGrand, C.~E.~Detar, S.~Datta and S.~A.~Gottlieb {\it et al.},
  Phys.\ Rev.\ D {\bf 64}, 054506 (2001)
  [hep-lat/0104002].

\bibitem{Zhou:2012sn} 
  R.~Zhou, S.~Gottlieb, J.~A.~Bailey, D.~Du, A.~X.~El-Khadra, R.~D.~Jain, A.~S.~Kronfeld and R.~S.~Van de Water {\it et al.},
  arXiv:1211.1390 [hep-lat].

\bibitem{Zhou:2011be} 
  R.~Zhou {\it et al.}  [Fermilab Lattice and MILC Collaborations],
  PoS LATTICE {\bf 2011}, 298 (2011)
  [arXiv:1111.0981 [hep-lat]].

\bibitem{Bouchard:2012tb} 
  C.~M.~Bouchard, G.~P.~Lepage, C.~J.~Monahan, H.~Na and J.~Shigemitsu,
  arXiv:1210.6992 [hep-lat].

\bibitem{Liu:2009dj} 
  Z.~Liu, S.~Meinel, A.~Hart, R.~R.~Horgan, E.~H.~Muller and M.~Wingate,
  PoS LAT {\bf 2009}, 242 (2009)
  [arXiv:0911.2370 [hep-lat]].

\bibitem{Liu:2011rra} 
  Z.~Liu, S.~Meinel, A.~Hart, R.~R.~Horgan, E.~H.~Muller and M.~Wingate,
  arXiv:1101.2726 [hep-ph].

\bibitem{Bazavov:2009bb} 
  A.~Bazavov, D.~Toussaint, C.~Bernard, J.~Laiho, C.~DeTar, L.~Levkova, M.~B.~Oktay and S.~Gottlieb {\it et al.},
  Rev.\ Mod.\ Phys.\  {\bf 82}, 1349 (2010)
  [arXiv:0903.3598 [hep-lat]].

\bibitem{Follana:2006rc} 
  E.~Follana {\it et al.}  [HPQCD and UKQCD Collaborations],
  Phys.\ Rev.\ D {\bf 75}, 054502 (2007)
  [hep-lat/0610092].

\bibitem{Sheikholeslami:1985ij} 
  B.~Sheikholeslami and R.~Wohlert,
  Nucl.\ Phys.\ B {\bf 259}, 572 (1985).

\bibitem{ElKhadra:1996mp} 
  A.~X.~El-Khadra, A.~S.~Kronfeld and P.~B.~Mackenzie,
  Phys.\ Rev.\ D {\bf 55}, 3933 (1997)
  [hep-lat/9604004].

\bibitem{Kronfeld:2000ck} 
  A.~S.~Kronfeld,
  Phys.\ Rev.\ D {\bf 62}, 014505 (2000)
  [hep-lat/0002008].

\bibitem{Oktay:2008ex} 
  M.~B.~Oktay and A.~S.~Kronfeld,
  Phys.\ Rev.\ D {\bf 78}, 014504 (2008)
  [arXiv:0803.0523 [hep-lat]].

\bibitem{Lepage:1992tx} 
  G.~P.~Lepage, L.~Magnea, C.~Nakhleh, U.~Magnea and K.~Hornbostel,
  Phys.\ Rev.\ D {\bf 46}, 4052 (1992)
  [hep-lat/9205007].

\bibitem{Horgan:2009ti} 
  R.~R.~Horgan, L.~Khomskii, S.~Meinel, M.~Wingate, K.~M.~Foley, G.~P.~Lepage, G.~M.~von Hippel and A.~Hart {\it et al.},
  Phys.\ Rev.\ D {\bf 80}, 074505 (2009)
  [arXiv:0906.0945 [hep-lat]].

\bibitem{Bernard:2007ma} 
  C.~Bernard, M.~Golterman and Y.~Shamir,
  Phys.\ Rev.\ D {\bf 77}, 074505 (2008)
  [arXiv:0712.2560 [hep-lat]].

\bibitem{Beneke:2001at} 
  M.~Beneke, T.~Feldmann and D.~Seidel,
  Nucl.\ Phys.\ B {\bf 612}, 25 (2001)
  [hep-ph/0106067].

\bibitem{Beneke:2004dp} 
  M.~Beneke, T.~.Feldmann and D.~Seidel,
  Eur.\ Phys.\ J.\ C {\bf 41}, 173 (2005)
  [hep-ph/0412400].

\bibitem{Chetyrkin:1996vx} 
  K.~G.~Chetyrkin, M.~Misiak and M.~Munz,
  Phys.\ Lett.\ B {\bf 400}, 206 (1997)
  [Erratum-ibid.\ B {\bf 425}, 414 (1998)]
  [hep-ph/9612313].

\bibitem{Bobeth:2010wg} 
  C.~Bobeth, G.~Hiller and D.~van Dyk,
  JHEP {\bf 1007}, 098 (2010)
  [arXiv:1006.5013 [hep-ph]].

\bibitem{Bailey:2008wp} 
  J.~A.~Bailey, C.~Bernard, C.~E.~DeTar, M.~Di Pierro, A.~X.~El-Khadra, R.~T.~Evans, E.~D.~Freeland and E.~Gamiz {\it et al.},
  Phys.\ Rev.\ D {\bf 79}, 054507 (2009)
  [arXiv:0811.3640 [hep-lat]].

\bibitem{Foley:2005ac} 
  J.~Foley, K.~Jimmy Juge, A.~O'Cais, M.~Peardon, S.~M.~Ryan and J.~-I.~Skullerud,
  Comput.\ Phys.\ Commun.\  {\bf 172}, 145 (2005)
  [hep-lat/0505023].

\bibitem{Lepage:2001ym} 
  G.~P.~Lepage, B.~Clark, C.~T.~H.~Davies, K.~Hornbostel, P.~B.~Mackenzie, C.~Morningstar and H.~Trottier,
  Nucl.\ Phys.\ Proc.\ Suppl.\  {\bf 106}, 12 (2002)
  [hep-lat/0110175].

\bibitem{Hornbostel:2011hu} 
  K.~Hornbostel, G.~P.~Lepage, C.~T.~H.~Davies, R.~J.~Dowdall, H.~Na and J.~Shigemitsu,
  Phys.\ Rev.\ D {\bf 85}, 031504 (2012)
  [arXiv:1111.1363 [hep-lat]].

\bibitem{Aubin:2007mc} 
  C.~Aubin and C.~Bernard,
  Phys.\ Rev.\ D {\bf 76}, 014002 (2007)
  [arXiv:0704.0795 [hep-lat]].

\bibitem{:2012uw} 
   A.~Bazavov,  {\it et al.} [The MILC Collaboration],
  arXiv:1212.4768 [hep-lat].

\bibitem{Durr:2010vn} 
  S.~Durr, Z.~Fodor, C.~Hoelbling, S.~D.~Katz, S.~Krieg, T.~Kurth, L.~Lellouch and T.~Lippert {\it et al.},
  Phys.\ Lett.\ B {\bf 701}, 265 (2011)
  [arXiv:1011.2403 [hep-lat]].

\bibitem{Durr:2010aw} 
  S.~Durr, Z.~Fodor, C.~Hoelbling, S.~D.~Katz, S.~Krieg, T.~Kurth, L.~Lellouch and T.~Lippert {\it et al.},
  JHEP {\bf 1108}, 148 (2011)
  [arXiv:1011.2711 [hep-lat]].

\bibitem{Bailey:2011bh} 
  J.~A.~Bailey {\it et al.}  [Fermilab and Lattice and MILC Collaborations],
  PoS LATTICE {\bf 2011}, 270 (2011)
  [arXiv:1111.5471 [hep-lat]].

\bibitem{Na:2010uf} 
  H.~Na, C.~T.~H.~Davies, E.~Follana, G.~P.~Lepage and J.~Shigemitsu,
  Phys.\ Rev.\ D {\bf 82}, 114506 (2010)
  [arXiv:1008.4562 [hep-lat]].

\end{thebibliography}
\end{document}